\documentclass[12pt,titlepage]{article}

\newcommand{\be}{\begin{equation}}
\newcommand{\ee}{\end{equation}}

\newcommand{\bi}{\begin{itemize}}
\newcommand{\ei}{\end{itemize}}

\newif\ifdraft
\drafttrue 

\usepackage[pdftex]{graphicx}
\DeclareGraphicsExtensions{.jpg,.pdf,.mps,.png}

\usepackage{caption}
\usepackage{helvet}
\usepackage[]{amsmath}

\usepackage{cite}
\let\citeleft=(
\let\citeright=)

\graphicspath{{images/}}

\begin{document}
\bibliographystyle{mrm}

\begin{center}
	\begin{Large}
		\begin{bf}
Reconstruction of Undersampled 3D Non-Cartesian Image-Based Navigators for Coronary MRA Using an Unrolled Deep Learning Model
\\ [0.1in]
		\end{bf}
	\end{Large}
\end{center}
\bigskip
\begin{center}
AUTHORS - Mario O. Malav\'e $^1$, Corey A. Baron $^2$, Srivathsan P. Koundinyan $^1$, Christopher M. Sandino $^1$, Frank Ong $^1$, Joseph Y. Cheng $^{1,3}$, and Dwight G. Nishimura $^1$

\end{center}
\vspace*{0.1in}
\noindent
1. Magnetic Resonance Systems Research Laboratory, Department of Electrical Engineering, Stanford University, Stanford, California \\ 
2. Department of Medical Biophysics, Western University, London, Ontario, Canada \\
3. Department of Radiology, Stanford University, Stanford, CA




\noindent
This work was supported in part by NSF Graduate Research Fellowship Program, NIH R01 HL127039, and GE Healthcare.

\newpage

\section*{Abstract}
\setlength{\parindent}{0in}
\textbf{Purpose:} To rapidly reconstruct undersampled 3D non-Cartesian image-based navigators (iNAVs) using an unrolled deep learning (DL) model for non-rigid motion correction in coronary magnetic resonance angiography (CMRA).
\\\textbf{Methods:} An unrolled network is trained to reconstruct beat-to-beat 3D iNAVs acquired as part of a CMRA sequence. The unrolled model incorporates a non-uniform FFT operator to perform the data consistency operation, and the regularization term is learned by a convolutional neural network (CNN) based on the proximal gradient descent algorithm. The training set includes 6,000 3D iNAVs acquired from 7 different subjects and 11 scans using a variable-density (VD) cones trajectory. For testing, 3D iNAVs from 4 additional subjects are reconstructed using the unrolled model. To validate reconstruction accuracy, global and localized motion estimates from DL model-based 3D iNAVs are compared with those extracted from 3D iNAVs reconstructed with $\textit{l}_{1}$-ESPIRiT. Then, the high-resolution coronary MRA images motion corrected with autofocusing using the $\textit{l}_{1}$-ESPIRiT and DL model-based 3D iNAVs are assessed for differences.
\\\textbf{Results:} 3D iNAVs reconstructed using the DL model-based approach and conventional $\textit{l}_{1}$-ESPIRiT generate similar global and localized motion estimates and provide equivalent coronary image quality. Reconstruction with the unrolled network completes in a fraction of the time compared to CPU and GPU implementations of $\textit{l}_{1}$-ESPIRiT (20x and 3x speed increases, respectively).
\\\textbf{Conclusion:} We have developed a deep neural network architecture to reconstruct undersampled 3D non-Cartesian VD cones iNAVs. Our approach decreases reconstruction time for 3D iNAVs, while preserving the accuracy of non-rigid motion information offered by them for correction.


\vspace{0.25in}
\setlength{\parindent}{0in}
{\bf Key words: convolutional neural networks, non-Cartesian, 3D cones trajectory, coronary MRA}
\newpage

\section*{Introduction}
We have previously developed an approach for free-breathing whole-heart coronary magnetic resonance angiography (CMRA) \cite{addy20173d, wu2013free} using an alternating-repetition time (ATR) balanced steady-state free precession (bSSFP) sequence \cite{leupold2006alternating}. The high-resolution imaging data is collected with a non-Cartesian 3D cones trajectory. For translational and non-rigid respiratory motion tracking, beat-to-beat 3D image-based navigators (iNAVs) of the heart are acquired  \cite{wu2013free,gurney2006design} using an accelerated variable-density (VD) 3D cones sampling technique \cite{addy20173d,luo2017nonrigid}. The reconstruction of 3D iNAVs using compressed sensing (i.e., $\textit{l}_{1}$-ESPIRiT \cite{uecker2014espirit,addy20173d}) is a time-consuming process, as each scan involves the collection of several hundred 3D iNAVs.

Deep learning (DL) has the potential to reduce reconstruction times for undersampled MRI data. Convolutional neural networks (CNNs) have recently become a powerful tool for image reconstruction. CNNs are popular due to their ease of use, accuracy, and fast inference time. CNN architectures generally operate in the image domain and are trained to minimize a specific loss function with respect to a “ground truth” image. One of the issues with a pure CNN architecture is the lack of incorporating physics specific to the application, leading to a black-box DL approach. Such an approach requires a very large number of training datasets, and can lead to issues with image quality as well as convergence during training. Through the use of a DL model-based architecture, more sophisticated techniques that incorporate CNNs with previous iterative reconstruction methods can provide improved accuracy while reducing the demand for training data and training time \cite{hammernik2018learning}.

Among the several DL model-based approaches that have been proposed, an unrolled network architecture \cite{diamond2017unrolled,cheng2018highly,cheng2019compressed} has emerged as a promising technique. Here, images are reconstructed by unfolding the proximal gradient descent (PGD) algorithm \cite{parikh2014proximal} and learning the regularization functions and coefficients. Prior studies leveraging unrolled networks have been limited to contexts involving Cartesian acquisitions. For reconstruction of non-Cartesian datasets, only image-to-image CNNs have been investigated \cite{hauptmann2019real}. In this work, we modify the unrolled model architecture to accommodate non-Cartesian 3D \textit{k}-space datasets. This would enable the rapid reconstruction of the undersampled 3D iNAV datasets acquired in our CMRA sequence.

\section*{Methods}
\subsection*{Imaging Data and 3D iNAV Acquisition}
Beat-to-beat undersampled 3D iNAVs are acquired as part of the CMRA sequence shown in Supporting Information Figure S1a. Specifically, free-breathing high resolution CMRA data (28x28x14 cm$^3$ FOV, 1.2 mm isotropic resolution, 500-600 total heartbeat scan time) is collected with a 3D cones trajectory using ATR-bSSFP \cite{wu2013free,addy20173d,malave2019whole}. The 3D iNAVs are acquired in the same volumetric region and after the segmented full-resolution acquisition by continuing the ATR-bSSFP sequence to maintain similar image contrast \cite{addy20173d}. The 3D cardiac datasets were acquired on a 1.5T GE Signa system with an 8-channel cardiac coil using VD trajectories consisting of 32 cone readouts, yielding an acceleration factor of 9 due to undersampling \cite{addy2015high} two different trajectories that were based on either sequential or phyllotaxis 3D iNAV designs \cite{malave2018stanford}. Details for both trajectories are shown in Supporting Information Figure S1b and S1c. 

To correct for rigid and nonrigid respiratory motion, an autofocusing technique is applied to the high resolution data. This motion compensation method requires accurate localized motion estimates, which are derived from the 3D iNAVs following their reconstruction with a computationally expensive iterative optimization approach (i.e., $\textit{l}_{1}$-ESPIRiT). The solution to this optimization problem, however, can be derived in an accelerated fashion using an unrolled DL model.

\subsection*{3D iNAV Reconstruction}
\subsubsection*{Iterative Algorithm and Unrolled Network Overview}
The unrolled network is based on PGD, which solves the following inverse problem with the image $x$, \textit{k}-space data $y$, encoding operator $A$, and regularization term $R(x)$:

\begin{equation} \label{eq:minObjfun}
\underset{x}{\text{minimize}}\;\frac{1}{2}||Ax-y||^2_2 + \lambda R(x)
\end{equation}

The solution, which is found using proximal gradient descent, iterates between the data consistency and proximal operator steps:

\begin{equation} \label{eq:updatefun}
x^{k+1} = P_R(x^k - \alpha A^T(Ax^k - y))
\end{equation}

The proximal operator of the regularization function is $P_R$, defined as:

\begin{equation}\label{eq:prox}
   P_R(v) = \underset{u}{\text{arg min }} (R(u) + \frac{1}{2 \lambda} ||u - v||_2^2)
\end{equation}

When using non-Cartesian data, the acquisition model $A$ incorporates the SENSE reconstruction \cite{pruessmann1999sense} operator $S$ (coil sensitivity maps computed using ESPIRiT \cite{uecker2014espirit}), and the non-uniform Fast Fourier Transform (NUFFT) operator, $F_{NUFFT}$. The regularization term is implicitly learned by replacing the proximal operator $P_R$ with a CNN to obtain the next iteration $x^{k+1}$. For $\textit{l}_{1}$-ESPIRiT, the regularization function is the $\textit{l}_{1}$-norm of the wavelet transform applied to the image, $x$. In this case, PGD simplifies to the iterative soft-shrinkage algorithm (ISTA) \cite{daubechies2004iterative}.

The data consistency step is important because it allows the model to incorporate the measured \textit{k}-space data in each iteration. There are two ways of implementing data consistency: hard-projection, and soft-projection. To apply hard-projection, the trajectory undersampling is performed using subsampling (i.e., not collecting certain data points from the fully sampled \textit{k}-space trajectory). After the acquisition, the uncollected data points are zero-filled. For each iteration of $\textit{l}_{1}$-ESPIRiT or the unrolled model, the Fourier transform of the 3D image is taken (after previous regularization, i.e., wavelet or CNN), and the data in the acquired \textit{k}-space trajectory locations are replaced with the original measured data while allowing for the zero-filled locations to update (Figure 1a). 

When using compressed sensing, VD sampling has been shown to work well \cite{lustig2007sparse}. Many different \textit{k}-space sampling techniques have been developed using VD Cartesian subsampling \cite{vasanawala2011practical,liu2014accelerated,prieto2015highly,cheng2015free}. For example, in Cheng et al. \cite{cheng2015free}, a Cartesian subsampling design called Variable-Density sampling and Radial view ordering (VDRad) is used which approximates VD spirals on the Cartesian grid. The current 3D iNAVs use a true VD non-Cartesian design \cite{addy20173d} instead of the subsampled Cartesian approach. The VD non-Cartesian cones design is not a subsampled version of a fully sampled trajectory; therefore, there are not any uncollected \textit{k}-space data points to zero-fill and utilize during the regularization and hard-projection steps. Thus, soft-projection is used for data consistency. In summary, for the Cartesian approach, hard-projection or soft-projection can be applied, but the VD non-Cartesian design is limited to soft-projection. To further illustrate the trajectory differences, a subsampled Cartesian trajectory, and a VD non-Cartesian trajectory are shown in Figure 1b. 

One key difference with the prior Cartesian model and proposed non-Cartesian model is the replacement of the data consistency step using the FFT with a gradient descent (GD) update step (i.e., soft-projection) using the NUFFT, which maintains consistency with the measured non-Cartesian \textit{k}-space data. The step size for the GD update step, $\alpha$, is initialized to 2 and left as a learnable parameter for the model to allow for improved training flexibility. Also, soft-projection for data consistency can potentially give improved results when the \textit{k}-space measurements are noisy, since a GD approach is used \cite{candes2006stable}. The non-Cartesian unrolled model is summarized below.

\underline{Unrolled Network Problem:} \\
$J_{ObjectiveFunction} = \frac{1}{2}||Ax-y||^2_2 + \lambda R(x)$ \\
$J_{DataConsistency} = \frac{1}{2}||Ax-y||^2_2$

$A$ = $F_{NUFFT}S$ (acquisition model = NUFFT operator, SENSE operator) \\
CNN = $\lambda$$R(x)$ (wavelet regularization is replaced with a CNN) \\
$y$ = input raw \textit{k}-space data \\
$x_{out}$ = output of the network \\
$\alpha$ = initialize to 2 (learnable parameter) \\
$k$ = index for each iteration ($N$ = number of iterations) \\
$x^{k}$ = output of the data consistency step \\
$x_{cnn}^k$ = output of the CNN

1. Initialize $x_{cnn}^0$ = $A^{T}y$ \\
2. Data consistency (soft projection/gradient update): \\
\hspace*{2cm}$x^{k+1} = x_{cnn}^k - \alpha\nabla(J_{DataConsistency}) = x_{cnn}^k - \alpha A^{T}(Ax_{cnn}^k-y)$ \\
3. CNN (regularization step): \\
\hspace*{2cm}$x_{cnn}^{k+1} = CNN(x^{k+1})$ (Eq. 2 with $P_R$ = CNN) \\
4. Repeat Steps $2 - 3$ for $(N-1)$ iterations \\
5. $x_{out} = x_{cnn}^{N}$

\subsubsection*{Neural Network Architecture}
The unrolled model architecture uses 4 gradient steps ($N$=4 iterations) consisting of 2 ($M$=2) residual network (ResNet) \cite{he2016deep} blocks/step. The hyperparameters were initially chosen to match \cite{cheng2018highly} and empirically tuned to ensure convergence given the memory constraints for the current application. The input into the network is the undersampled 3D complex \textit{k}-space data, \textit{k}-space coordinates (to generate the NUFFT operator), and the respective coil sensitivity maps for each channel (for the SENSE operator). The ground truth used for training is the reconstructed image when using $\textit{l}_{1}$-ESPIRiT. Each gradient step begins with data consistency which uses the forward and transpose acquisition model $A$ and $A^{T}$ to apply soft projection where $x^0$ is initialized as $A^{T}y$. The complex image is then separated into 2 channels consisting of the real and imaginary components. Next, the network uses M ResNet blocks comprised of two 3D convolutional layers with a kernel size of 3x3x3 and filter depth of 64. Also, each convolutional layer is preceded by a ReLU pre-activation layer as recommended in He et al. \cite{he2016identity}. An additional layer is added to the end of each unrolled step which outputs 2 channels for the real and imaginary parts of the \textit{k}-space data and uses a linear activation to preserve the sign of the data. The final layer is also added to a skip connection from the input of the first ResNet block to accelerate training convergence. For previous Cartesian approaches \cite{cheng2018highly,cheng2019compressed}, circular convolutions were used to handle the “wrap-around” coherent aliasing artifacts. However, while the FFT causes periodic boundary conditions in the image domain, here the NUFFT operation does not because of image cropping and zero-padding that it uses, and the noise-like aliasing properties of undersampled 3D cones \cite{gurney2006design}. Accordingly, zero-padded convolutions (i.e., non-circular convolutions) were applied in each convolutional layer. The gradient step block is then repeated 3 (i.e., $N-1$) more times for a total of $N$=4 iterations. Also, the network is trained using the complex $l_1$ loss and a batch size of 1 (stochastic gradient descent). A graphical representation of the unrolled architecture is shown in Figure 2a and 2b. In Figure 2b, the data consistency for the prior and proposed models are shown using a hard-projection and soft-projection respectively. 

\subsection*{Training Data}
The training set includes a total of seven subjects. Four of the seven subjects were scanned with both the sequential and phyllotaxis based 3D iNAVs acquired each heartbeat which gives a total of eleven scans. Five of the eleven scans were acquired with the trajectories rotated by the golden angle between each heartbeat (Supporting Information Figure S1d) to vary the aliasing artifacts, which serves as a form of data augmentation to improve the performance and ability of the model to generalize. Each scan collected data for 500-600 heartbeats, thus yielding approximately 6,000 total iNAV datasets used for training. The ground truth images were reconstructed with $\textit{l}_{1}$-ESPIRiT using the Berkeley Advanced Reconstruction Toolbox (BART) toolbox \cite{uecker2015berkeley}. The $\textit{l}_{1}$-ESPIRiT reconstruction parameters were empirically determined using 50 iterations, step size of 1e-6, and wavelet regularization ($\lambda$= 0.05). To alleviate the aliasing artifacts from objects outside of the FOV during training, the ground truth datasets were reconstructed at (2$\times$FOV$_x$, 1.25$\times$FOV$_y$, 1.5$\times$FOV$_z$); i.e., (128$_x$, 80$_y$, 48$_z$) with a native matrix size of (64$_x$, 64$_y$, 32$_z$). The reconstruction was run on two different Linux systems with 2.20 GHz Xeon E5-2650 v4 CPU, 512 GB RAM with 48 total cores, and a 3.70 GHz Intel i7-8700K CPU, 64 GB RAM with 12 total cores. The reconstruction was also performed on two different GPUs using an NVIDIA Titan XP with 12 GBs of GDDR5X memory, and an NVIDIA Titan RTX GPU with 24 GBs of ultra-fast GDDR6 memory.

\subsubsection*{Computation}
When using the NUFFT operator, there is an increase in computation compared to the standard FFT. The NUFFT operator requires additional steps involving density compensation, convolution with a gridding kernel (Kaiser-Bessel) \cite{beatty2005rapid}, sampling on the Cartesian grid, and an apodization correction. This can introduce challenges when reconstructing undersampled datasets which require larger matrix sizes to alleviate aliasing artifact. When using the NUFFT operator with the unrolled model, the increased computation increases the memory requirements for training, thus limiting the matrix size. Additionally, once the data points are interpolated onto the Cartesian grid, a radix-2 Cooley-Tukey FFT \cite{norton1987parallelization} was used to decrease memory usage. To satisfy memory requirements, the proposed model architecture was implemented in Python using TensorFlow and trained on an NVIDIA Titan RTX GPU with 24 GBs of ultra-fast GDDR6 memory.

\subsection*{Motion Correction with 3D iNAVs}
To correct for respiratory motion in the high resolution data, 3D global and localized motion estimates are calculated \cite{ingle2014nonrigid,luo2017nonrigid,addy20173d} using both the $\textit{l}_{1}$-ESPIRiT and model-reconstructed 3D iNAVs. A technique similar to the previous state-of-the-art method in Luo et al. \cite{luo2017nonrigid} was applied for motion-estimate calculation when using the 3D iNAVs:

1. The motion estimates are generated by selecting a region of interest (ROI) mask that covers the heart in the axial, sagittal and coronal planes. Then, a reference 3D iNAV time frame (heartbeat) is determined using mutual information \cite{studholme1999overlap,pluim2003mutual} and the similarity matrix approach. 

2. Global translational motion estimates are calculated by minimizing the mean-squared difference cost function between the current 3D iNAV frame with the previously determined reference 3D iNAV.

3. To take advantage of the localized spatial information from the 3D iNAVs, residual 3D displacement fields are also calculated, using the MATLAB Imaging Processing Toolbox (The Mathworks, Natick, MA), relative to the reference 3D iNAV (after aligning the 3D iNAVs using the global motion estimates) and used to determine five unique spatial regions (or bins) of localized residual motion. The bins are obtained with k-means clustering (minimizing the $l_2$-norm distance metric), and using the displacement field estimates (in the selected ROI) as the features. 

4. The mean of all the features within each calculated bin (plus the global motion estimate) is then used as the residual motion estimate for the bin. 

We then apply a linear phase modulation term generated using the motion estimates for each heartbeat in \textit{k}-space to generate a bank of six 3D motion-compensated reconstructions from one global motion estimate, and five residual localized motion estimates (all five applied on top of the global estimates). For the final non-rigid autofocused image, the reconstruction is performed on a pixel-by-pixel basis by choosing the pixel from the bank of 3D motion-compensated reconstructions that minimizes the gradient entropy value at each pixel \cite{ingle2014nonrigid,addy20173d}.

\subsection*{Inference and Testing}
Four additional subjects were scanned with the previously mentioned navigator designs (2 sequential-based, 1 rotated sequential-based, and 1 rotated phyllotaxis-based) to test the generalization of the unrolled model. The motion information of the 3D iNAVs when using $\textit{l}_{1}$-ESPIRiT and the unrolled model is then assessed by examining the motion estimates, autofocusing outcomes, and right coronary artery (RCA) and left coronary artery (LCA) images. The motion estimate similarity between the $\textit{l}_{1}$-ESPIRiT and DL model-based 3D iNAVs is determined by calculating the correlation coefficients of the left/right (L/R), anterior/posterior (A/P), and superior/inferior (S/I) motion estimates for the global and five spatial bins. Furthermore, the autofocusing outcomes were analyzed by computing the ``autofocusing histograms" for each volunteer. The autofocusing histograms show the occurrence of each selected bin that minimized the gradient entropy for each pixel in the final high-resolution image. Also, oblique reformatted maximum intensity projection (MIP) images of the RCA and LCA are shown with cross-sectional views of the vessels before and after motion correction when using the $\textit{l}_{1}$-ESPIRiT and DL model-based 3D iNAVs for autofocusing.

\section*{Results}
\subsection*{3D iNAV Reconstructions}
All four subject (test dataset) 3D iNAV inputs (gridded images), outputs, and ground truths (after $\textit{l}_{1}$-ESPIRiT) are shown in (Figure 3) with their respective axial, sagittal, and coronal slices. An example output for each iteration during training is shown in Supporting Information Figure S2. For each iteration, image depiction is improved by denoising the image and enhancing structure which mimics multiple iterations for $\textit{l}_{1}$-ESPIRiT. When employing the trained architecture, the undersampled cardiac images (compared to the outcomes from gridding) retained structural features as a result of the denoising/smoothening operation. More specifically, the aliasing artifacts arising from undersampling a cones trajectory were effectively reduced after evaluation by the network. The training was run for 40 epochs which took a total of approximately 400 hours. Inference time for the proposed architecture is approximately 0.5 seconds per 3D iNAV on GPU (Titan RTX), while $\textit{l}_{1}$-ESPIRiT (using BART) requires approximately 10 seconds on CPU (Xeon E5-2650 v4, and Intel i7-8700K) and 1.5 seconds on GPU (Titan RTX) as shown in Table 1. All four subject datasets have a total of 500-600 3D iNAVs.

\subsection*{Motion Estimates}
The global motion estimates for the first 100 heartbeats of the four different subjects are shown in Supporting Information Figure S3 for all three directions (A/P, L/R, S/I). The scatter plots for $\textit{l}_{1}$-ESPIRiT vs. Model and the correlation coefficients (R) are shown in Figure 4. Also, the measurements are normalized by subtracting the mean S/I, A/P, and L/R displacements similar to \cite{ingle2014nonrigid} when comparing the estimates. The additional correlation coefficients (including the global and five spatial bins) are shown in Supporting Information Table S1. The R values for all four subjects show a strong positive correlation of the global motion estimates between the $\textit{l}_{1}$-ESPIRiT and DL model-based 3D iNAVs, thus indicating similar motion estimates in all directions (Supporting Information Figure S3).

\subsection*{Autofocusing Histograms}
The autofocusing histograms for all four subjects are shown in Supporting Information Figure S4. For subjects 2 and 3 (Supporting Information Figure S4b and S4c respectively), the global bin is the most selected by the autofocusing algorithm which demonstrates that there was less residual motion beyond the rigid-body translational motion. For subjects 1 and 4 (Supporting Information Figure S4a and S4d), bins four and five are the most selected respectively which shows that there was additional residual motion which minimized the gradient entropy metric. Subjects 1 and 2 histograms lack noticeable difference, and subjects 3 and 4 show minor differences in bin 2. These minor differences are further investigated in the high-resolution images to verify the effects on the coronary image quality.

\subsection*{Motion-Corrected Images}
The 3D autofocused images using both 3D iNAV reconstruction schemes are shown for all four subject scans. The total time for autofocusing reconstruction takes approximately 8 mins (2 mins to calculate the global translations, 5 mins to calculate the displacement fields and perform k-means clustering, and 1 min to calculate gradient entropy minimization of each pixel).  In Figure 5, the right coronary artery (RCA) is shown before after motion correction with cross-sectional views demonstrating the improvements when using the $\textit{l}_{1}$-ESPIRiT and DL model-based 3D iNAVs. The RCA images for all four subjects show nearly identical vessel sharpness. Also, in Figure 6, the left coronary artery (LCA) is shown with the corresponding cross-sectional views. Similar to the RCA, the LCA sharpness increased after motion correction and maintained similar and comparable improvements when using $\textit{l}_{1}$-ESPIRiT and DL model-based 3D iNAVs for motion correction.

\section*{Discussion and Future Work}
We have shown that the proposed non-Cartesian unrolled network architecture generates similar 3D iNAV reconstruction results in a fraction of the time, with a 20x and 3x speed increase for CPU and GPU respectively, and leads to high correlations of the derived global and localized motion estimates compared to $\textit{l}_{1}$-ESPIRiT 3D iNAV reconstructions. The most computationally expensive part of the training involves the NUFFT operation. The NUFFT is used to apply the data consistency (soft projection) step which can require more calculations for the gradients during training, thus increasing training time and the GPU memory requirements. Similar issues arise during the $\textit{l}_{1}$-ESPIRiT iterative reconstruction. To account for this, BART uses the Toeplitz method \cite{baron2018rapid} for the compressed sensing reconstruction. However, the Toeplitz method requires a matrix size that is twice as large as the nominal matrix (i.e., twice the FOV in all three dimensions) to encompass the support of the object in the image domain. This helps to alleviate aliasing errors that can arise from objects outside of the FOV when using the Toeplitz method. Due to the computation benefits, further investigation may be warranted by using a Toeplitz-based NUFFT for the data consistency (soft projection) steps even though the 2x oversampling constraints are not required when using a normal NUFFT operator. Benefiting from using the Toeplitz-based NUFFT operator would depend on the amount of undersampling, trajectory type, and native matrix size \cite{ou2017nufft}.

When using the proposed non-Cartesian unrolled model, there are improvements and limitations to address. For this implementation, the complex data was separated into 2 channels without any noticeable training issues, but architectures that can process the full complex data channel may prove beneficial for better generalization and potentially lead to faster training since fewer filter weights are used. In Virtue et al. \cite{virtue2017better}, this was done by using complex activation layers that attenuate the magnitude based on the input phase, essentially acting as a “complex ReLU” activation function. Also, currently an $l_1$ loss is used, but other approaches such as using generative adversarial networks (GANs) \cite{goodfellow2014generative} have the ability to learn better loss functions that take into account diagnostic image quality \cite{kulkarni2016reconnet,mardani2017deep,quan2018compressed}. This would have the potential for improving the “perceived” image quality for clinicians which an $l_1$ or $l_2$ loss may not fully quantify. Furthermore, when collecting the 3D iNAV datasets, the fully sampled data is not obtained due to the finite acquisition window after collecting the segmented high-resolution data within one heartbeat; thus, the ground truth was obtained through the compressed sensing reconstruction of VD cones. Even though the ground truth is biased towards the compressed sensing reconstruction, we have shown that the unrolled network was able to successfully reconstruct the 3D iNAVs with improved image quality compared to gridding and allow for similar motion estimates compared to using compressed sensing. Finally, we have primarily trained the unrolled model with 3D VD cones cardiac datasets with 4.4 mm resolutions. Fortunately, due to flexibility of the model architecture, the training dataset for the current application is not limited to these specific undersampled 3D cardiac iNAVs. Thus, to help further generalization of the model, datasets acquired using different trajectories, resolutions, and anatomies can be added to the current training dataset for retraining (fine-tuning). When training with higher resolutions, the aliasing artifacts may become more severe and require larger matrix sizes to avoid aliasing from objects outside of the FOV during the data consistency step in the model. Despite these limitations, the model was able to successfully reconstruct 3D VD cones datasets and similarly improve RCA and LCA sharpness (when used as 3D iNAVs for motion correction) for the current application. Further analysis of different resolutions and anatomies may be warranted to test the limitations of the unrolled model.

In the current work, autofocusing \cite{addy20173d,luo2017nonrigid,ingle2014nonrigid} and binning with localized motion correction were investigated when using the proposed DL model-based 3D iNAV reconstruction approach. Further techniques for motion correction involve using a deep-learning framework \cite{lv2018respiratory,pawar2018moconet,lossau2019motion} to obtain the motion estimates or motion-corrected images directly from the \textit{k}-space data can potentially replace autofocusing but would likely require a larger training set. This could allow for an end-to-end model using the proposed unrolled PGD architecture for 3D iNAV reconstruction and an additional model which takes in the 3D iNAVs with the motion-corrupted high-resolution \textit{k}-space data and outputs the motion-corrected images. This framework has the potential for a substantial reduction in reconstruction time. Even if the end-to-end model produces less optimal motion-corrected images compared to previous techniques, the model can be used as a tool for quickly validating scan quality. Then, if further improvements are required, reconstruction can be applied using the standard more time-consuming $\textit{l}_{1}$-ESPIRiT and autofocusing techniques.

Future work includes implementing the non-Cartesian unrolled model for higher resolution data. Current limitations include the 24 GBs of memory on the NVIDIA RTX GPU which limits the potential matrix size used for training. To solve this problem, filter depth, number of iterations, and ResNet blocks can be decreased but may reduce the ability for proper generalization of the model. As previously mentioned, a Toeplitz-based NUFFT can also be used but would lead to a 2x FOV oversampling requirement. Also, multi-GPU training can be implemented to allow for larger matrix sizes but would increase training time due to introduced overhead. The technique that currently has the most potential and feasibility involves the bandpass approach \cite{cheng2018highly} which separates \textit{k}-space into blocks and requires segmenting \textit{k}-space and windowing the segments for stitching the final results of the output denoised image blocks. Training a non-Cartesian high-resolution unrolled model has many challenges and constraints, but there are many options for solving the memory and computation problems. 
\section*{Conclusion}
A deep neural network architecture was developed for the reconstruction of undersampled VD cones 3D iNAVs acquired during a CMRA sequence. The unrolled network architecture was designed to solve the PGD reconstruction problem and for the reconstruction of undersampled non-Cartesian datasets. It was shown that the reconstruction of 3D iNAVs using the DL model-based reconstruction compared to using $\textit{l}_{1}$-ESPIRiT can be performed in a fraction of the time ($1/20^{th}$ on CPU and $1/3^{rd}$ on GPU) while generating similar motion estimates and, after motion correction of the high-resolution data, equivalent RCA and LCA image quality.

\begin{figure}[h]
  \centering
  \includegraphics[width=\textwidth]{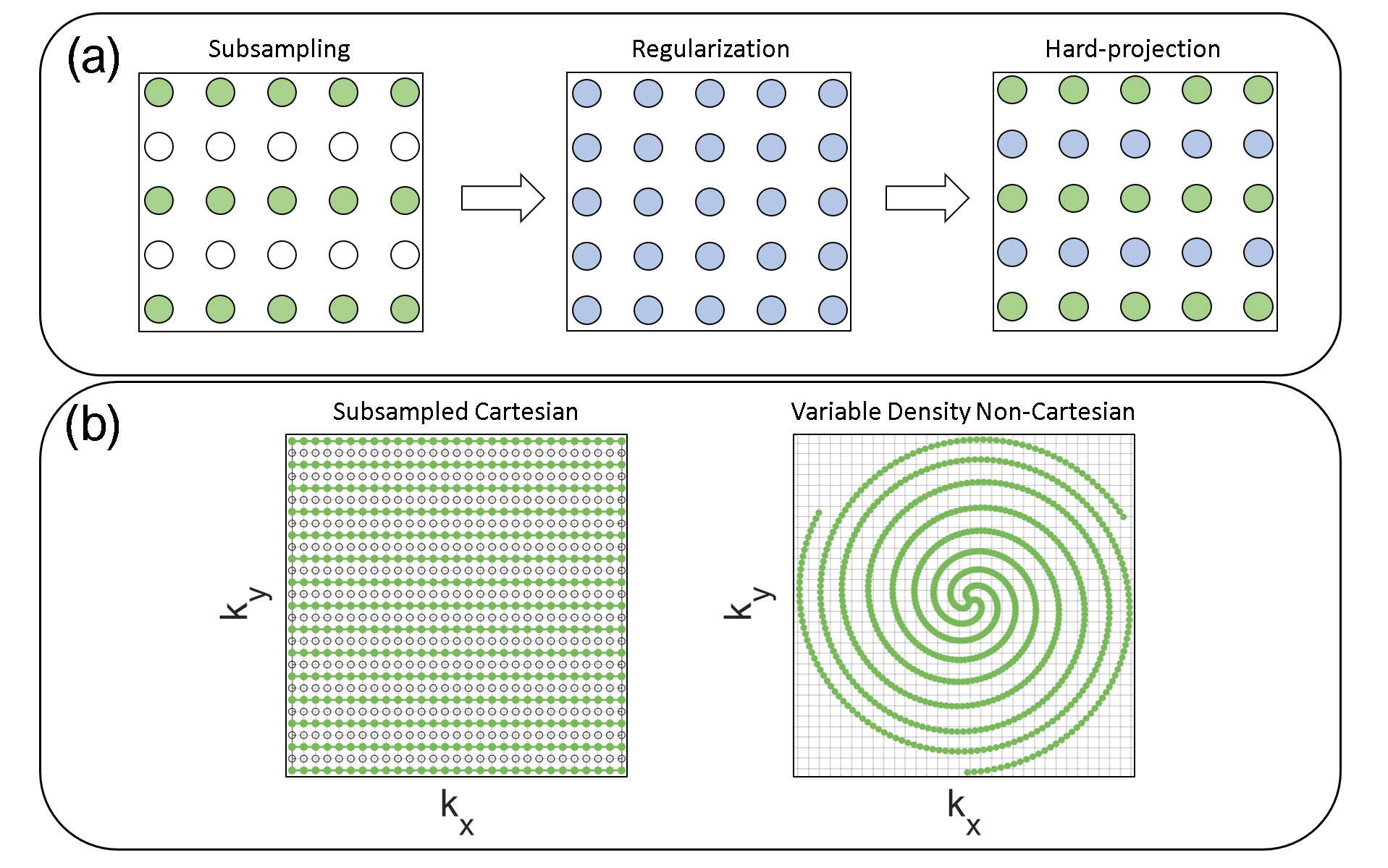}
  \caption[]
    {(a) A subsampled trajectory is shown where the green circles and white circles represent measured and zero-filled data respectively. After previous regularization using wavelets or a CNN, the Fourier transform of the 3D image is taken and the original \textit{k}-space trajectory locations are replaced with the measured data by performing a hard-projection. (b) The Cartesian trajectory (left) uses subsampling while the non-Cartesian trajectory (right) uses a variable-density (VD) design to achieve undersampling.
    }
\end{figure}

\begin{figure}
  \centering
  \includegraphics[width=\textwidth]{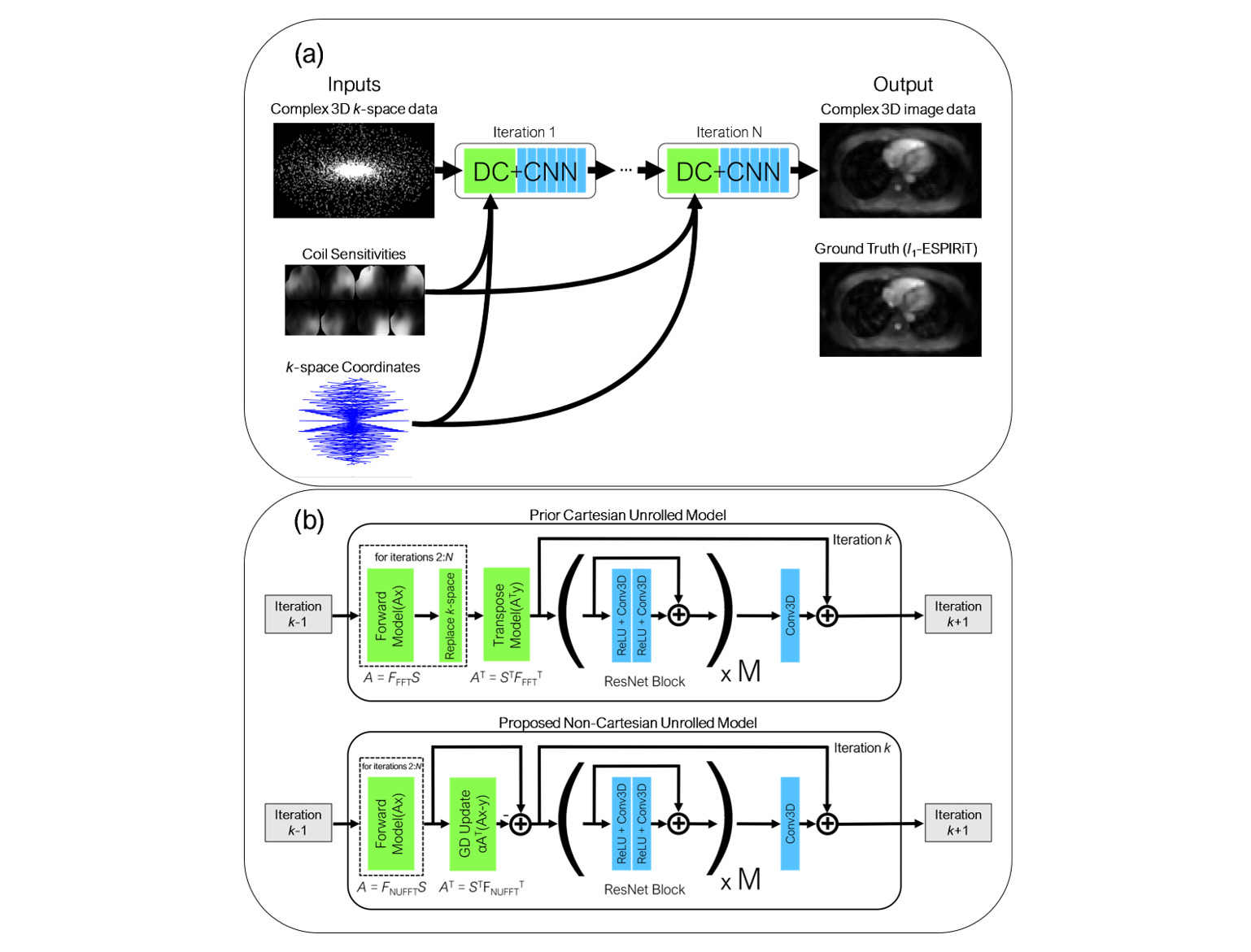}
  \caption[]
    {(a) The input into the network is the 3D \textit{k}-space data, \textit{k}-space coordinates (to generate the NUFFT operator, $F_{NUFFT}$), and the coil sensitivity maps for each channel (for the SENSE reconstruction operator $S$). The ground truth is the $\textit{l}_{1}$-ESPIRiT reconstruction of the input \textit{k}-space data. Each iteration consists of a data consistency (DC) and CNN block. (b) The architecture uses $N$=4 iterations (gradient steps) consisting of $M$=2 ResNet blocks/step. One key difference between the prior Cartesian model (top) and the proposed model (bottom) is the replacement of the data consistency step using an FFT with a gradient-descent (GD) update step (i.e., soft-projection) using the NUFFT, which maintains consistency with the measured non-Cartesian \textit{k}-space data. The first iteration using hard-projection (top) and soft-projection (bottom) only apply the transpose model ($A^Ty$) and gradient update step, respectively, followed by the CNN. The remaining iterations (i.e., 2 to $N$) follow the architectures shown in Figure 2b.
    }
\end{figure}

\begin{figure}
  \centering
  \includegraphics[width=\textwidth]{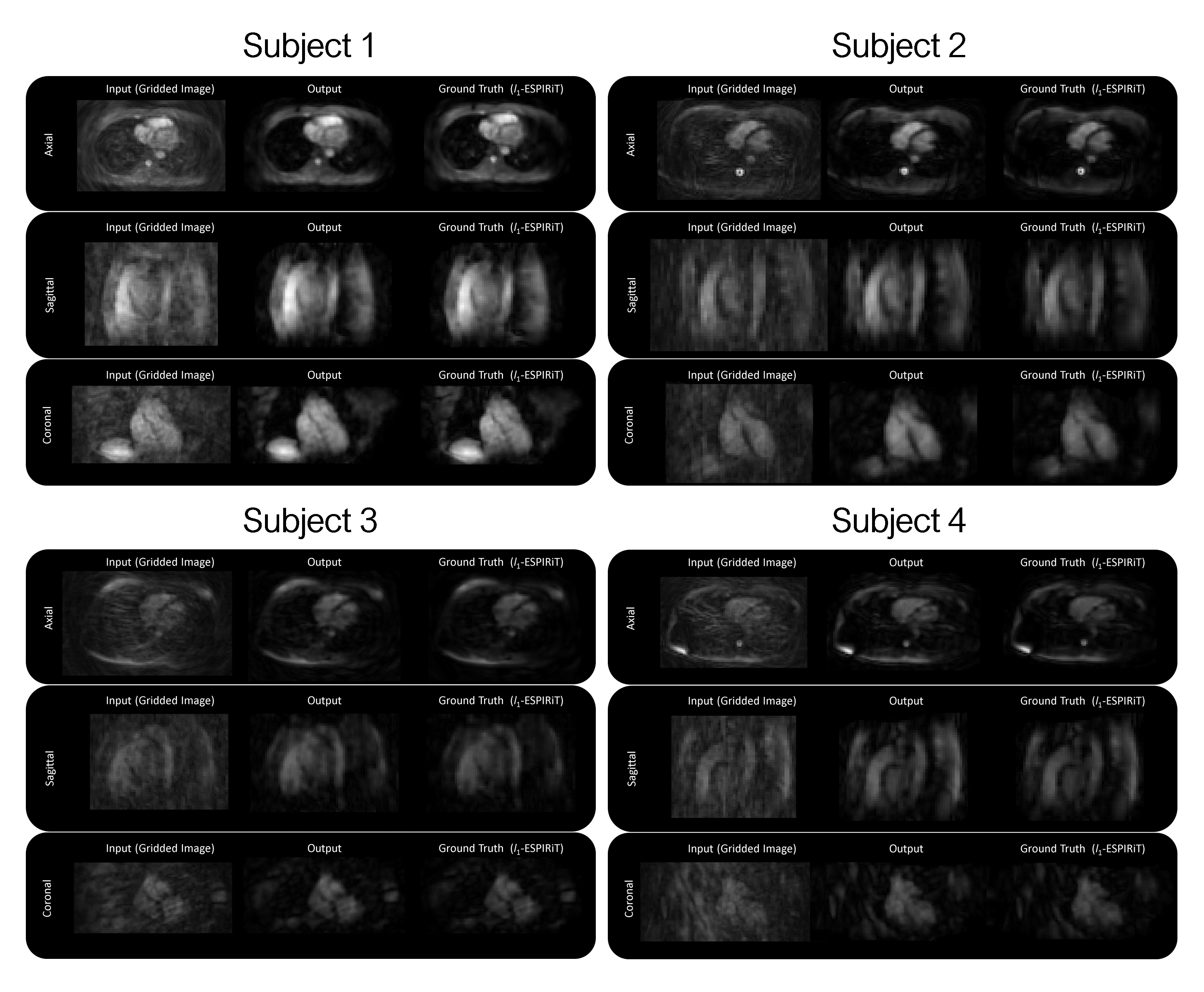}
  \caption[]
    {The axial, sagittal, and coronal slices are shown from one heartbeat. 3D iNAV inputs (gridded images using the NUFFT operator), outputs, and ground truths ($\textit{l}_{1}$-ESPIRiT) are shown respectively for four subject datasets.
    }
\end{figure}

\begin{figure}
  \centering
  \includegraphics[width=\textwidth]{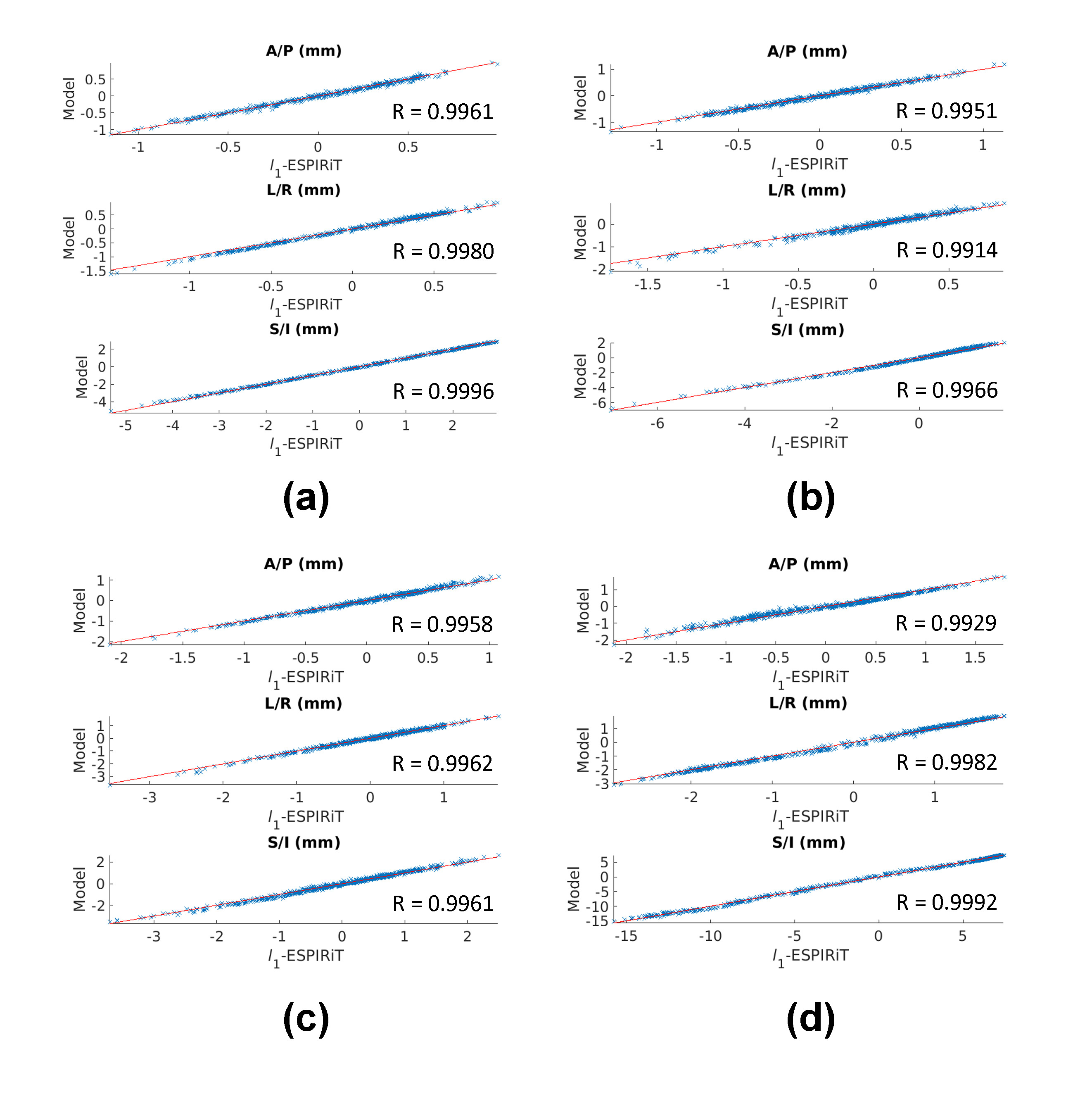}
  \caption[]
    {The global motion estimates for subjects 1-4 (a-d) are shown in three different scatter plots (A/P, L/R, S/I) with the corresponding correlation coefficients (R) for $\textit{l}_{1}$-ESPIRiT versus the model.
    }
\end{figure}

\begin{figure}
  \centering
  \includegraphics[width=\textwidth]{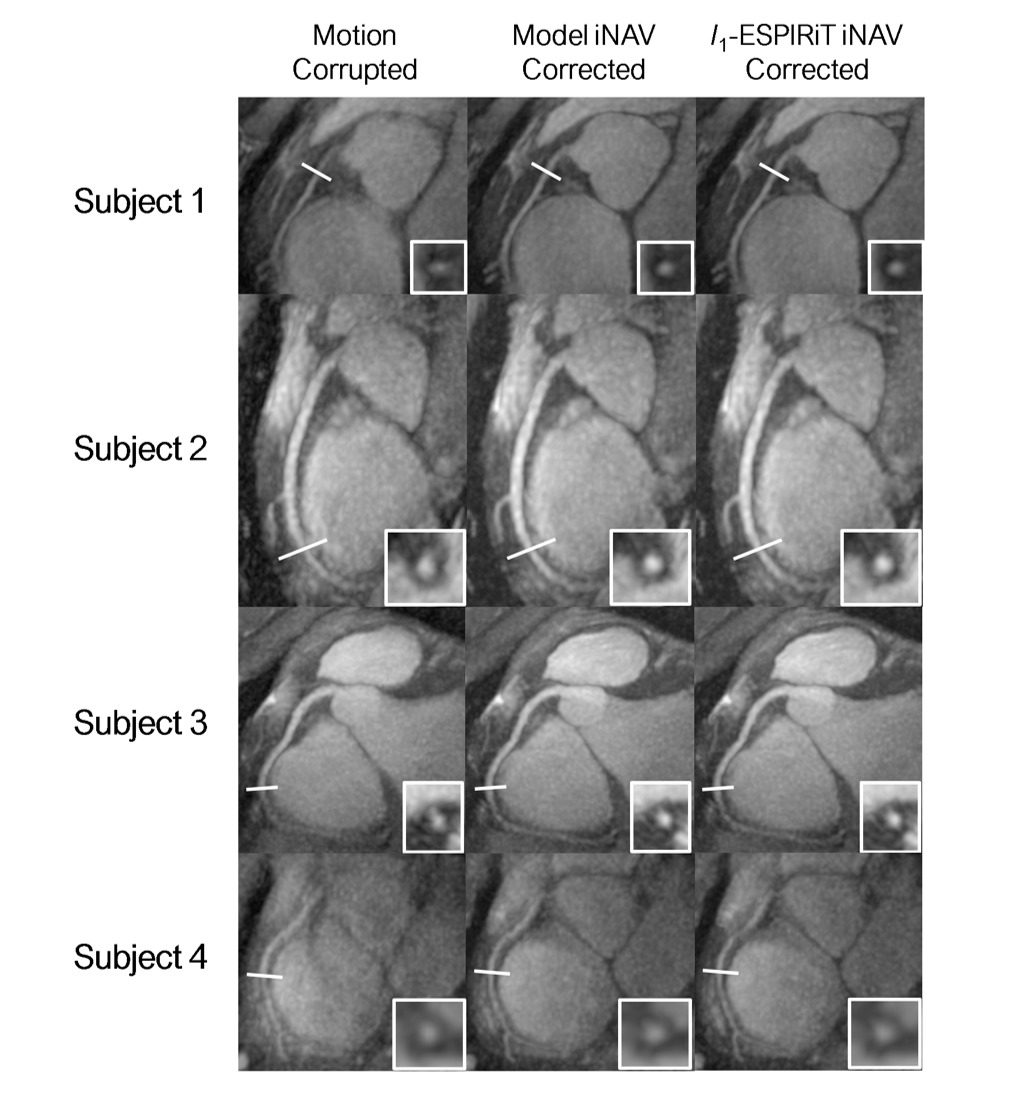}
  \caption[]
    {(a) Reformatted maximum intensity projection images of the RCA for four healthy volunteers are shown before and after motion correction. The cross-sectional views demonstrate similar improvements in the distal regions of the RCA when using both the $\textit{l}_{1}$-ESPIRiT and DL model-based 3D iNAVs for motion correction.
    }
\end{figure}

\begin{figure}
  \centering
  \includegraphics[width=\textwidth]{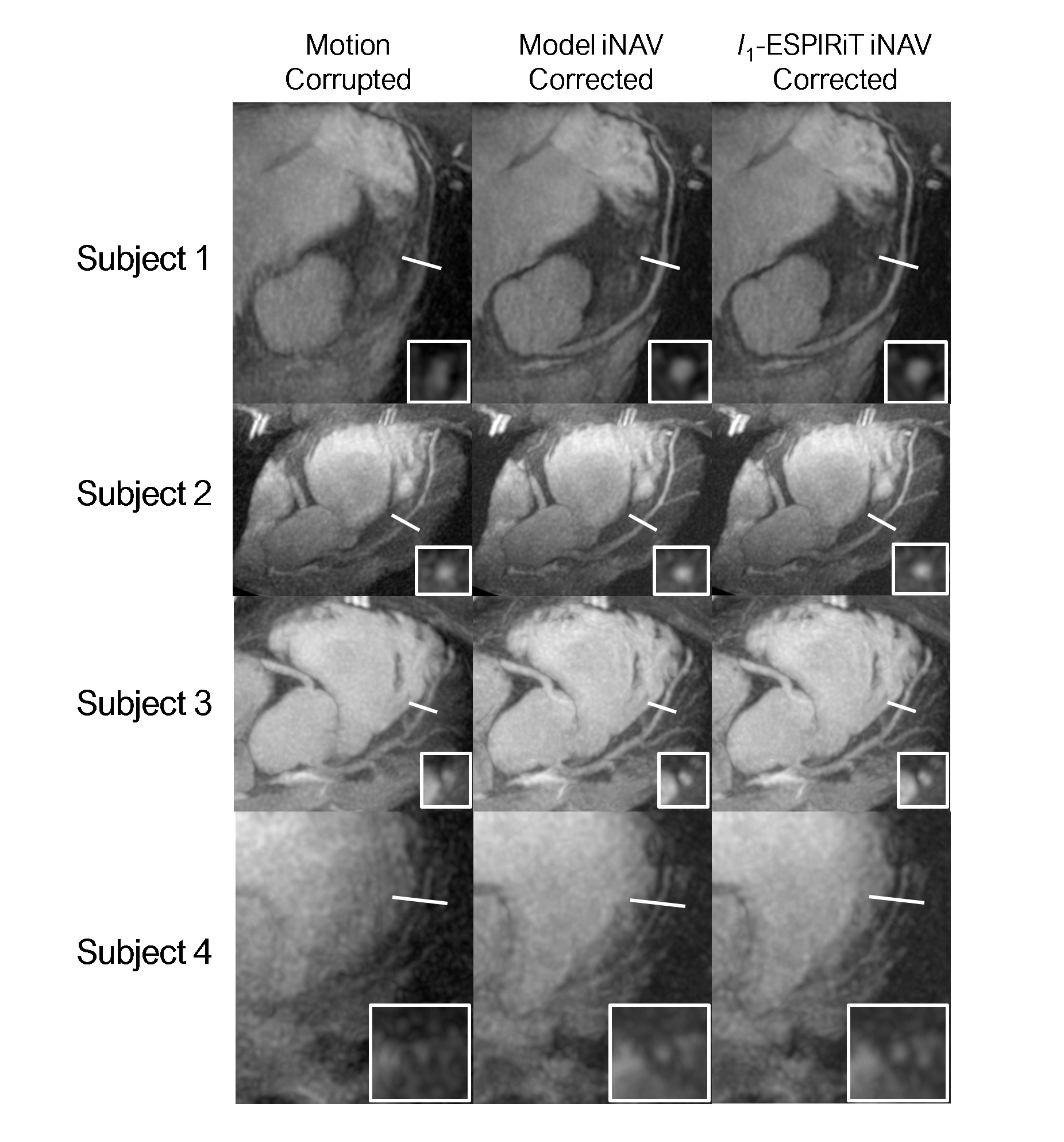}
  \caption[]
    {(a) Reformatted maximum intensity projection images of the LCA for four healthy volunteers are shown before and after motion correction with the corresponding cross-sectional views. The LCA sharpness increased in the medial and distal regions after motion correction and exhibited similar improvements when using $\textit{l}_{1}$-ESPIRiT and DL model-based 3D iNAVs for motion correction.
    }
\end{figure}

\begin{table}
  \centering
  \includegraphics[width=\textwidth]{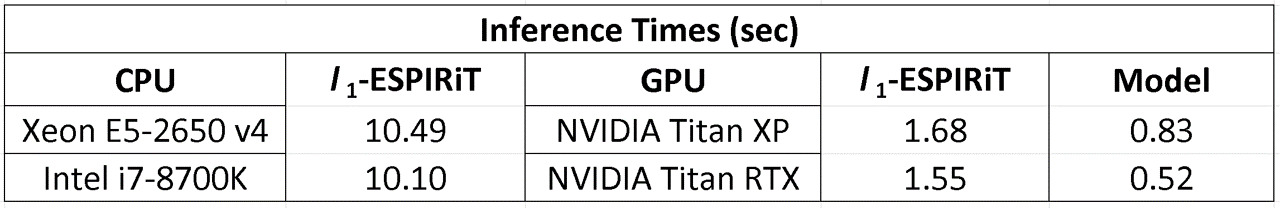}
  \caption[]
    {Reconstruction times for a single 3D iNAV on different CPU and GPU devices using $\textit{l}_{1}$-ESPIRiT and DL model.
    }
\end{table}

\newpage
\bibliography{refs_momalave}

\setcounter{figure}{0}
\begin{figure}[h]
  \centering
  \includegraphics[width=\textwidth]{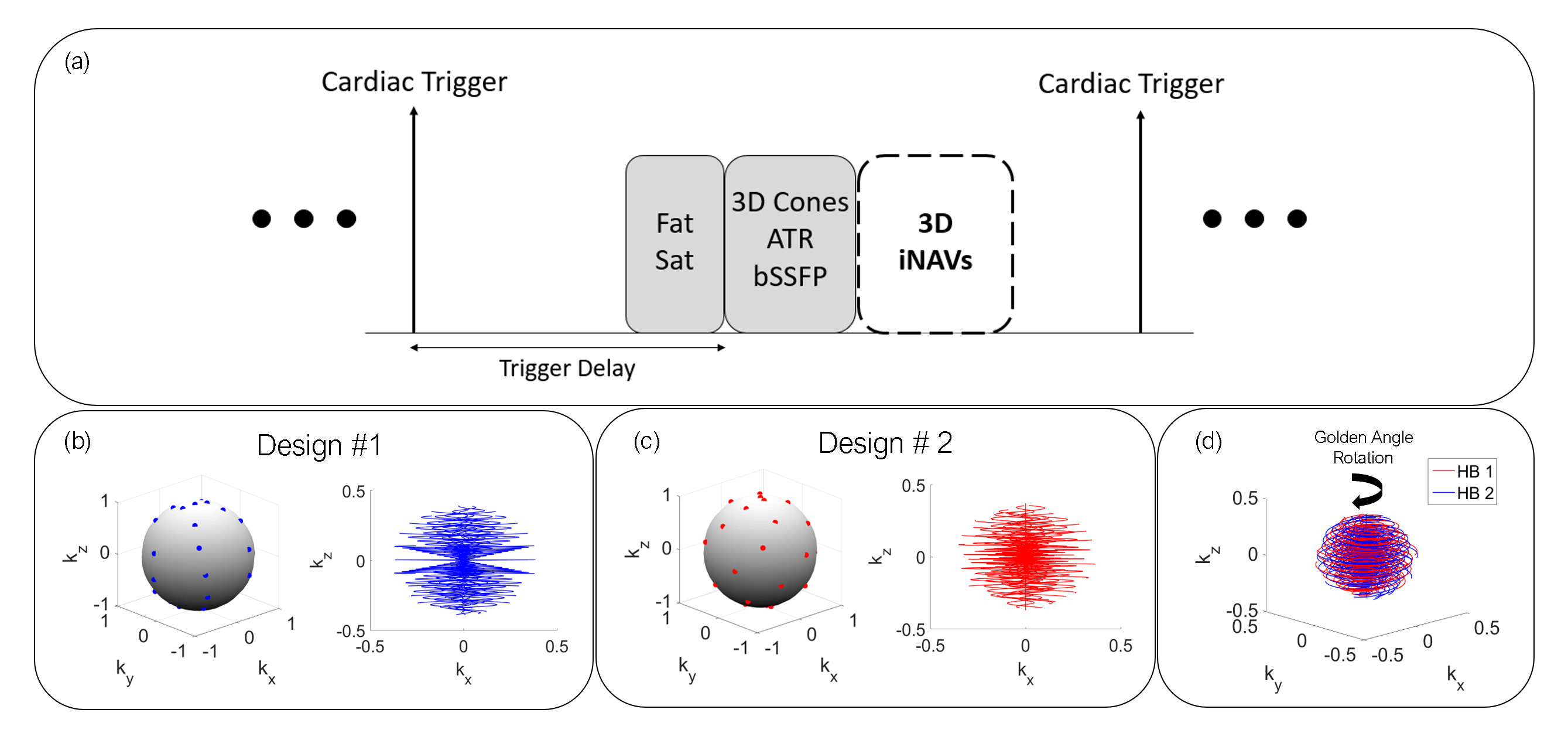}
  \renewcommand{\figurename}{Supporting Figure}
  \renewcommand{\thefigure}{S\arabic{figure}}
  \caption[]
    {(a) For the free-breathing CMRA acquisition scheme, the 3D iNAVs are collected every heartbeat following the fat saturation and imaging data acquisition as shown in the timing diagram. The 3D iNAVs are acquired using a variable-density, undersampled 3D cones trajectory. (b) The first design uses a sequential-based acquisition with multiple readouts (and uniform azimuthal rotations) within each conical surface. (c) The second design employs a phyllotaxis scheme with unique conical surfaces and golden angle azimuthal rotations. The blue and red points on the unit sphere represent the polar angles for each corresponding cone readout. (d) In addition, some of the datasets rotate the two trajectory designs between heartbeats by the golden angle to help the model further generalize during training.
    }
\end{figure}

\begin{figure}[h]
  \centering
  \includegraphics[width=\textwidth]{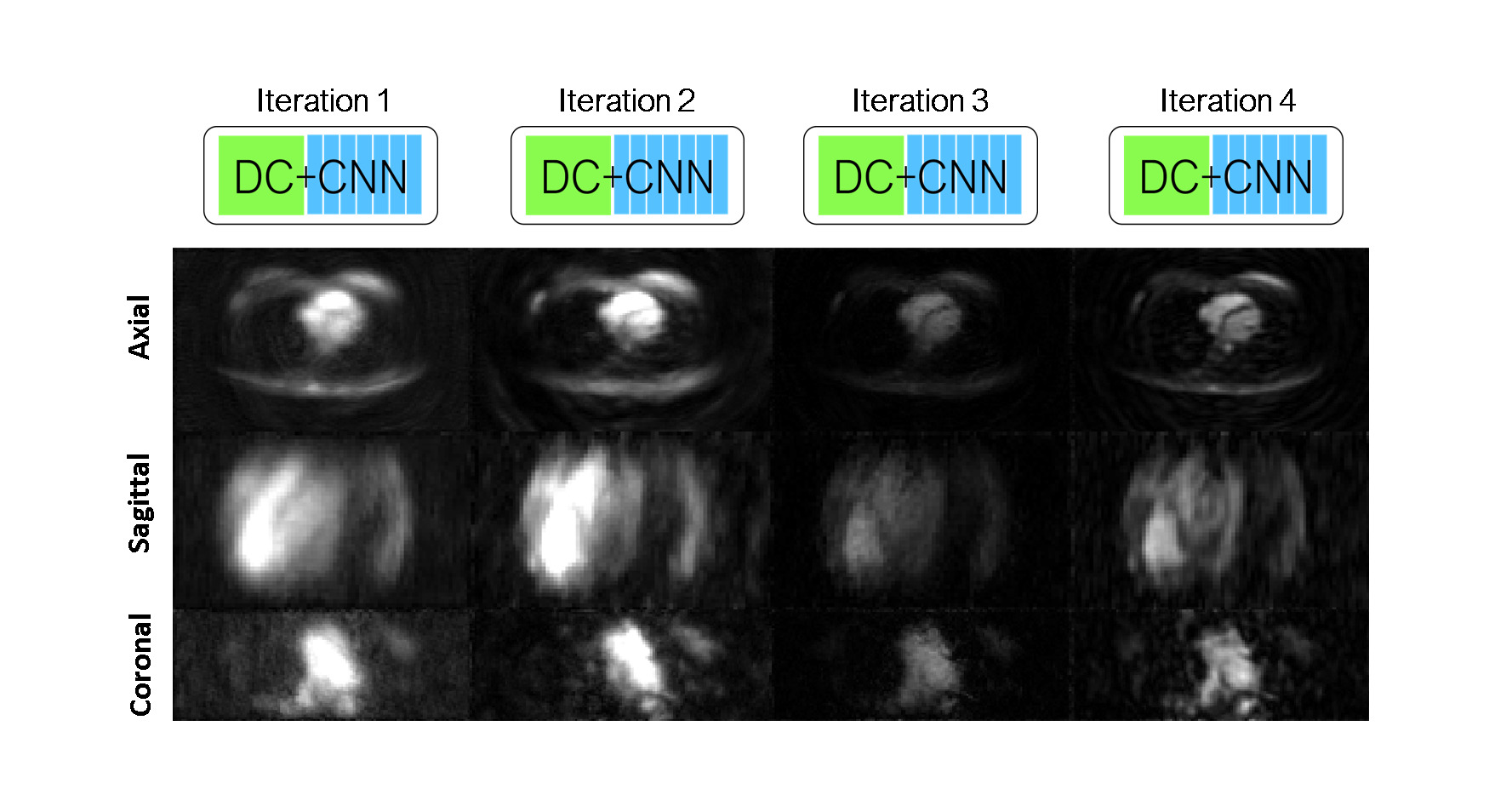}
  \renewcommand{\figurename}{Supporting Figure}
  \renewcommand{\thefigure}{S\arabic{figure}}
  \caption[]
    {(a) The outputs of each iteration in the unrolled model during training are shown for one example dataset. The respective outputs for each of the 4 iterations (gradient steps) highlight the behavior of each different blocks in the model. For each iteration, image depiction is improved by enhancing the structures throughout the axial, sagittal, and coronal slices. 
    }
\end{figure}

\begin{figure}[h]
  \centering
  \includegraphics[width=\textwidth]{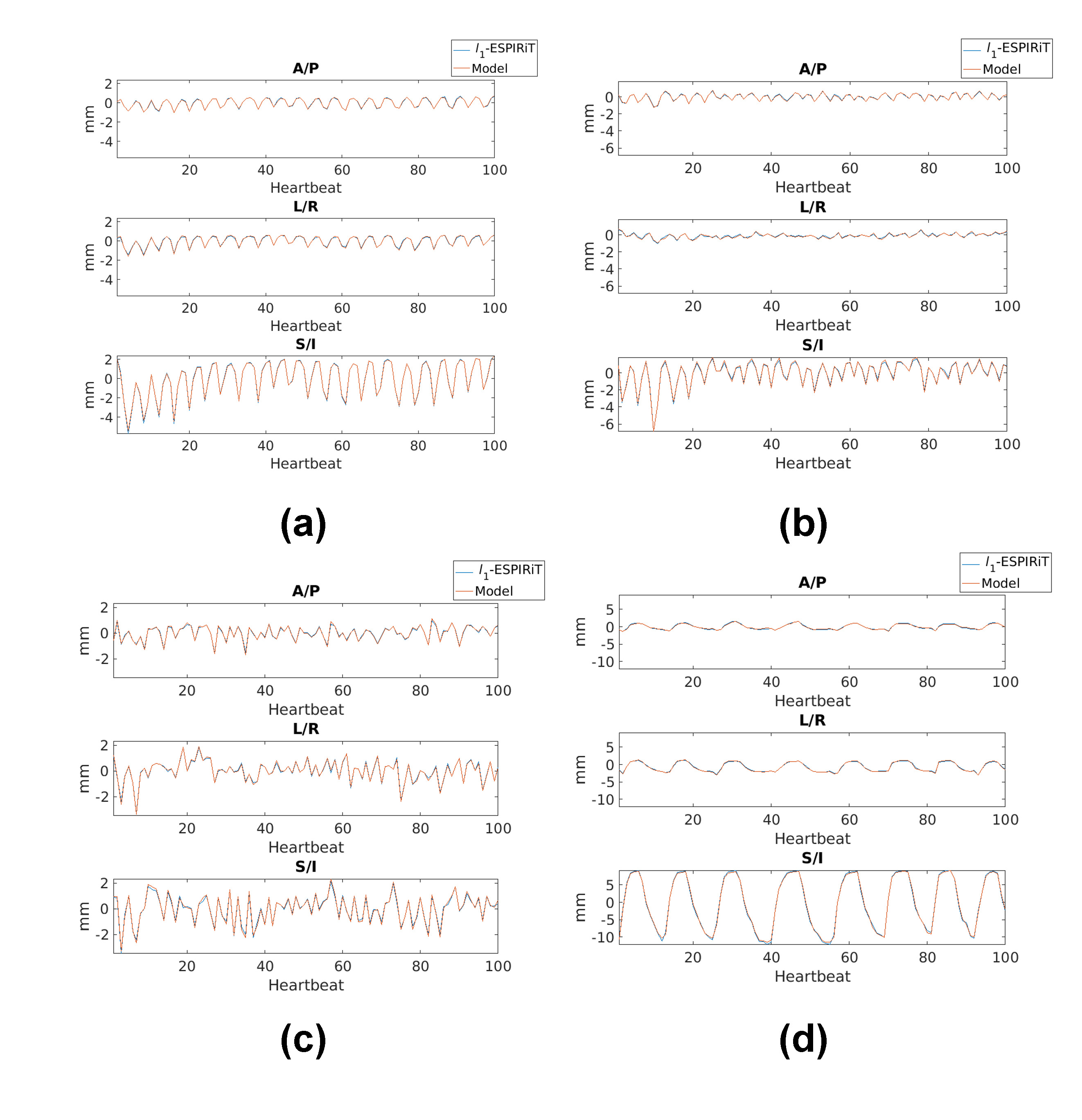}
  \renewcommand{\figurename}{Supporting Figure}
  \renewcommand{\thefigure}{S\arabic{figure}}
  \caption[]
    {The global motion estimates (first 100 heartbeats) generated from $\textit{l}_{1}$-ESPIRiT, and the DL model-based 3D iNAVs. The plots show how the motion estimates extracted from the $\textit{l}_{1}$-ESPIRiT and DL model-based 3D iNAVs track similar motion in all directions for all four subjects (a-d).
    }
\end{figure}

\begin{figure}[h]
  \centering
  \includegraphics[width=\textwidth]{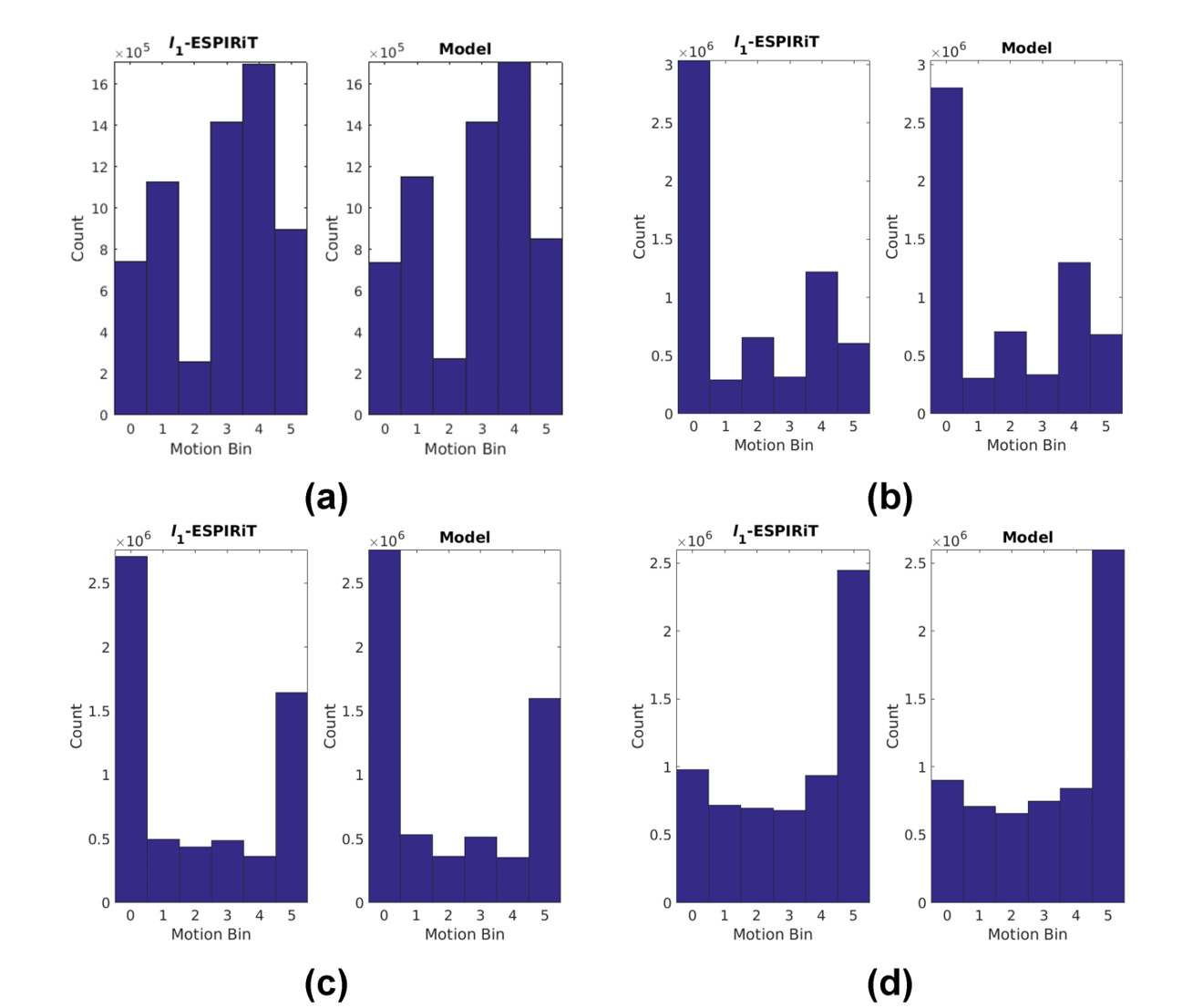}
  \renewcommand{\figurename}{Supporting Figure}
  \renewcommand{\thefigure}{S\arabic{figure}}
  \caption[]
    {The histograms generated from the outcomes of the autofocusing algorithm when using $\textit{l}_{1}$-ESPIRiT, and the DL model-based 3D iNAVs for subjects 1-4 (a-d). The histograms show the global and residual motion bins (0-5), respectively. For subjects 2 and 3 (b,c), the global bin is the most selected by autofocusing which shows that there was less residual motion beyond the rigid-body translational motion. For subjects 1 and 4 (a,d), bins four and five are the most selected, demonstrating that there was additional residual motion beyond translational.
    }
\end{figure}

\begin{figure}[h]
  \centering
  \includegraphics[width=\textwidth]{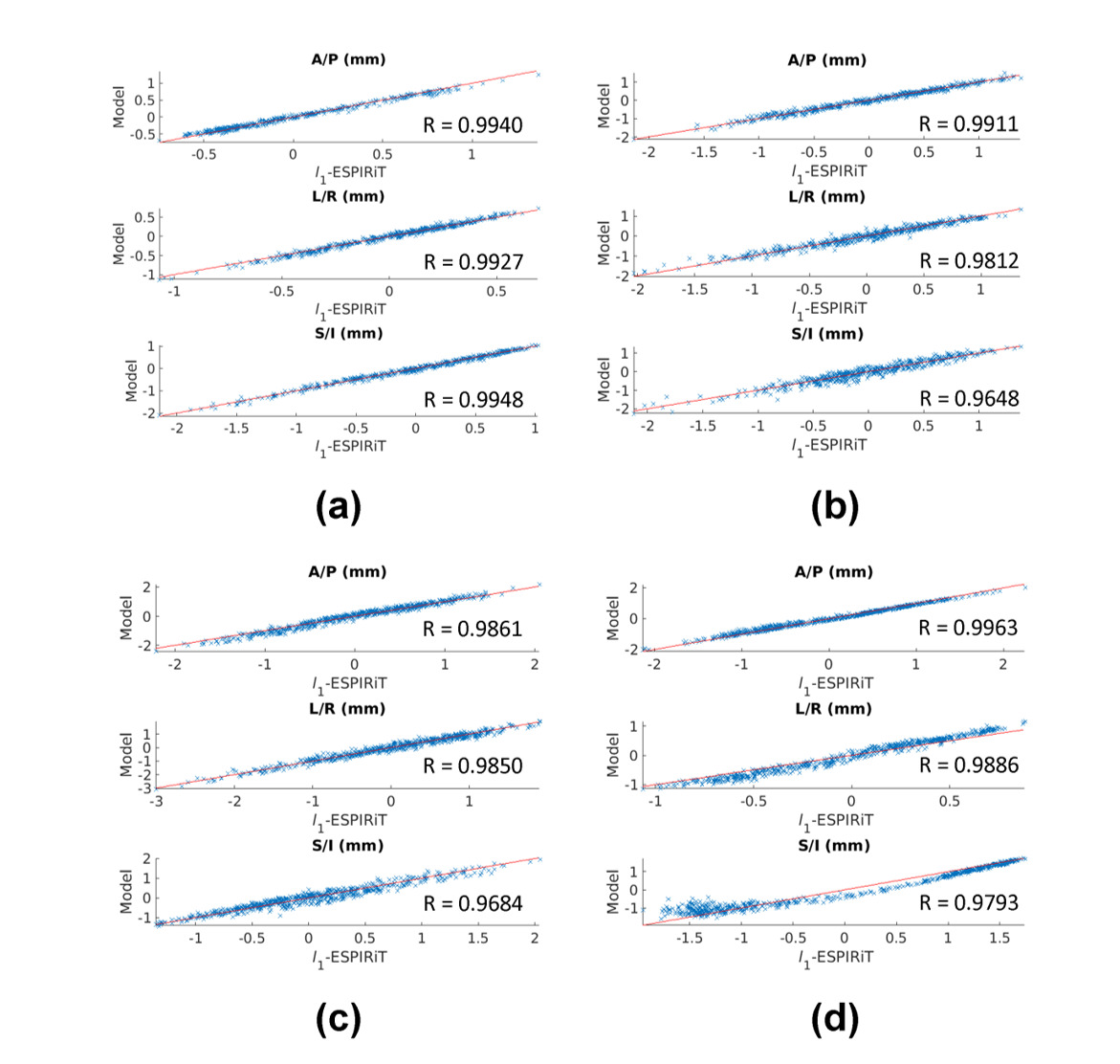}
  \renewcommand{\figurename}{Supporting Figure}
  \renewcommand{\thefigure}{S\arabic{figure}}
  \caption[]
    {(a) The most occurring residual motion estimate scatter plots (according to the autofocusing histogram bins from Supporting Information Figure S4) and correlation coefficients (R) for all four subjects (a-d) generated from $\textit{l}_{1}$-ESPIRiT, and the DL model-based 3D iNAVs. For subject 1 and 2, bin 4 is shown, and for subjects 3 and 4, bin 5 is shown. The scatter plots show slightly less correlation compared to the global estimates (Figure 4) which may partly be attributed to minor interpolation differences between the $\textit{l}_{1}$-ESPIRiT, and the DL model-based reconstructions.
    }
\end{figure}

\begin{figure}[ht]
  \centering
  \includegraphics[width=\textwidth]{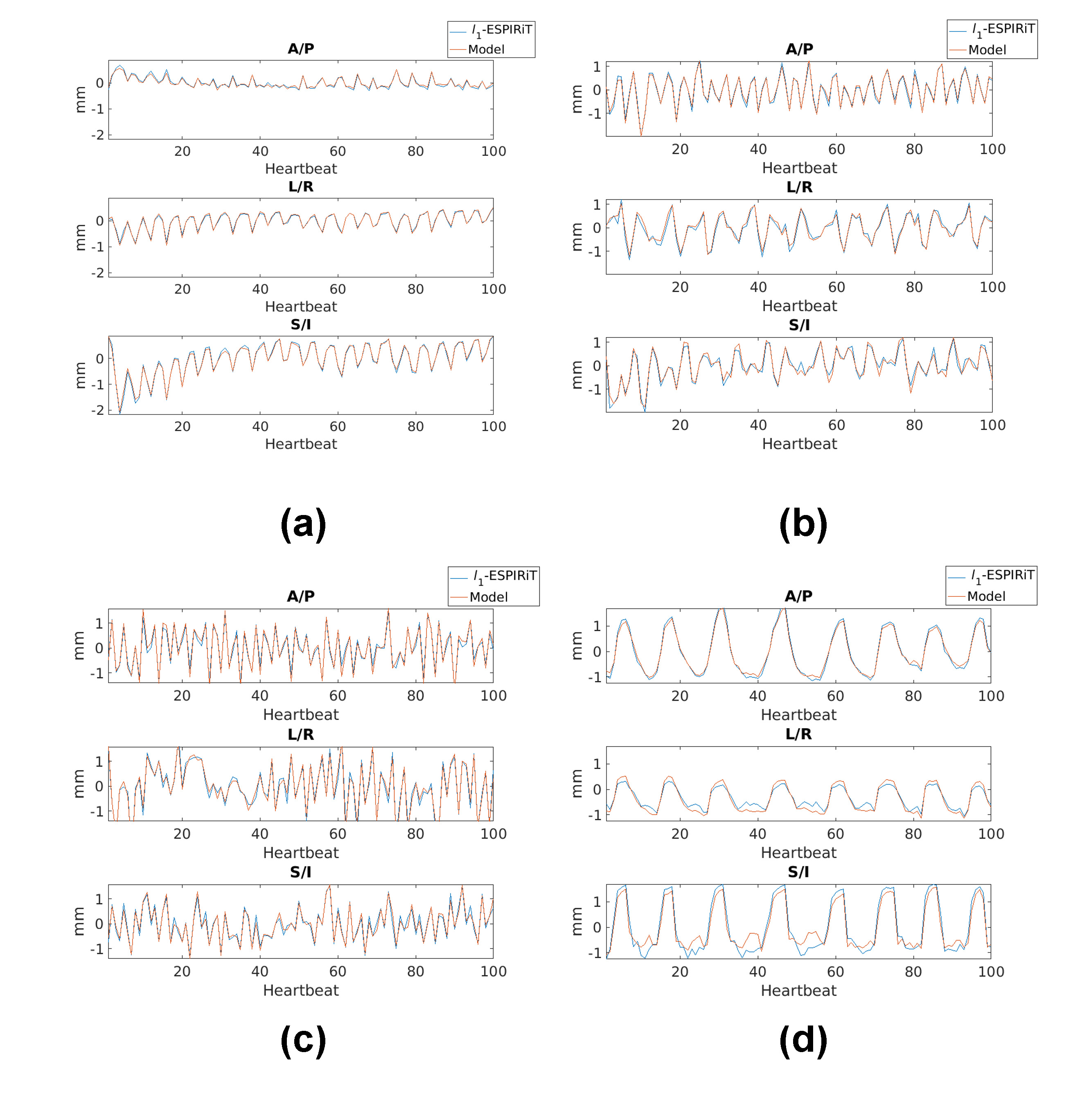}
  \renewcommand{\figurename}{Supporting Figure}
  \renewcommand{\thefigure}{S\arabic{figure}}
  \caption[]
    {(a) The most occurring residual motion estimates (first 100 heartbeats) for all four subjects (a-d) generated from $\textit{l}_{1}$-ESPIRiT, and the DL model-based 3D iNAVs. The corresponding motion bin estimates from Supporting Information Figure S5 are shown. The residual motion estimates (A/P, L/R, S/I) allow for residual motion correction which the global translations do not fully capture.
    }
\end{figure}

\begin{table}[h]
  \centering
  \includegraphics[width=\textwidth]{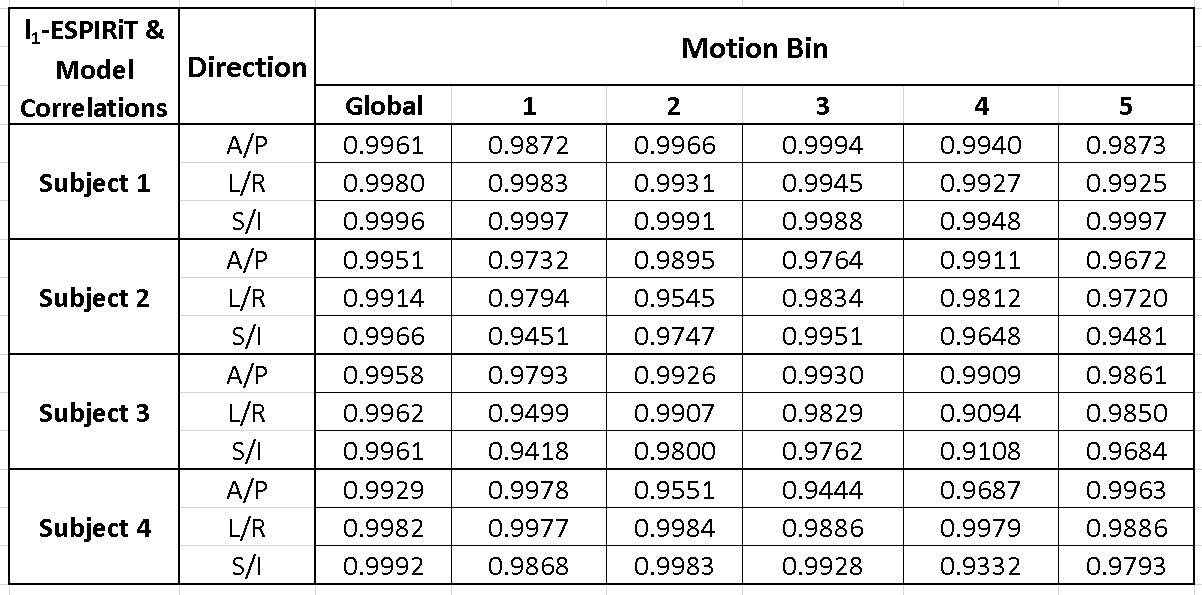}
  \caption*
    {Supporting Information Table S1: The correlation coefficients between motion estimates (in A/P, L/R, and S/I) obtained from $\textit{l}_{1}$-ESPIRiT and the DL model-based 3D iNAVs for the global and five spatial bins. These motion estimates are used to generate a bank of six 3D motion-compensated reconstructions (from one global motion estimate, and five residual localized motion estimates) used as candidates for the autofocusing algorithm.
    }
\end{table}




\end{document}